\documentstyle[12pt,epsf]{article}
\begin{document}
\input{psfig}
\begin{titlepage}
\begin{center}
\hfill IP/BBSR/2000-35\\
\vskip .2in

{\Large \bf Noncommutative $N=2$ Strings}
\vskip 0.5in
{Alok Kumar, Aalok Misra and
Kamal Lochan Panigrahi\\
Institute of physics,\\
Bhubaneswar 751 005, India\\email: kumar, aalok, kamal@iopb.res.in}
\vskip 0.5 true in
\end{center}

\begin{abstract}
We analyze open and mixed sector tree-level amplitudes of 
$N=2$ strings in a space-time with (2,2) signature, 
in the presence of constant $B$ field. The expected topological 
nature of string amplitudes in the open sector is shown to impose 
nontrivial constraints on the corresponding noncommutative
field theory. In the mixed sector, we first compute 
a 3-point function and show that the corresponding field
theory is written in terms of a generalized 
*-product. We also analyze a 4-point function 
($A_{oooc}$) of the mixed sector in $\Theta \rightarrow \infty$ limit.

\end{abstract}
\end{titlepage}
\newpage
 
String Theories with $N=2$ worldsheet supersymmetry \cite{ADEMO}
have been an 
important area of research\cite{OV,KET,MART,MAR,BERK}
 due to their connection with 
self-dual gravity and Yang-Mills. Such string theories live on a
K\"{a}hler manifold with (2,2) signature and their tree amplitudes
have a `magical' \cite{OV} property that the  
$n$-point functions are either local, or zero ( for $n\geq 4$),
thus having no `effective'  propagating degrees of freedom. These
theories being intimately related to M(atrix) and F
theories\cite{KET,MART}, are hence considered important from the point of
view of obtaining the nonperturbative
fundamental theory as well.

In this paper we study $N=2$ strings in constant 
NS-NS antisymmetric tensor ($B$) background,
in view of interesting developments  in noncommutative string 
theory\cite{DOUG,SW,SOLITONS, OMNCSOL}. In this regard, 
we have also been motivated by the fact that 
noncommutative $N=2$ strings are expected to have interesting
implications in possible generalizations of 
M(atrix) and F theories to include noncommutativity.

It is known that antisymmetric 
tensor backgrounds can be incorporated in $N=2$ superspace 
formalism using chiral and twisted-chiral superfields\cite{HULL}.
In this manner, one has an $N=2$ worldsheet supersymmetry
without having a K\"{a}hler metric \cite{HULL}. 
One now obtains a noncommutative complex manifold as the target space geometry.

There are two main highlights 
of the $N=2$ noncommutative field theory obtained in this paper. 
The first is a nontrivial constraint, satisfied by the noncommutative
theory, originating from the requirement 
of the absence of poles in the 4-point amplitude in the 
open sector of the $N=2$ strings for nonzero $B$. 
The second is in the mixed sector. Here the noncommutative field
theory involves a generalized *-product\cite{GAROUSI,MEHEN}, and
$B$ explicitly appears with the open string
metric in two (left/right) linear combinations for contracting 
the target space indices of the
open-string fields. We finally analyze a 4-point mixed sector 
amplitude in the extreme noncommutative limit. 
Nontriviality in the computation of the tree-level 
string amplitudes in the mixed sector, 
stems from the fact that due to
the absence of $z\rightarrow{\bar z}$-symmetry in the presence of
$B$, one can not use the  generalized Koba-Nielson integrals of 
\cite{OV,MAR,KAW} which are relevant when the domain of integration is
the full complex plane.

One can consider $N=2$ string action (in presence of
$B$), written in an $N=1$ superspace notation in \cite{HULL}:
\begin{equation}
\label{eq:1}
S=\int d^2x \int d\theta_L d\theta_R\biggl[g_{IJ}D^\alpha X^I
D_\alpha X^J
+B_{IJ}D^\alpha X^I(\gamma^5D)_\alpha X^J\biggr],
\end{equation}
where the superspace field $X^I$ ($I\equiv1, 2,
{\bar 1}, {\bar 2}$) represents $X, Y, {\bar X}, {\bar Y}$, and  
$\alpha\equiv L,R$. The closed string metric and the 
antisymmetric background fields are denoted by 
$g_{IJ}$ and $B_{IJ}$ respectively.

Field equations, boundary conditions and canonical 
commutation relations, including those for fermions
in $N=2$ case turn out to be   
similar to the ones written in 
\cite{CH,SW}. Since the vacuum energy of the bosonic and 
fermionic oscillators remains same as for $B=0$\cite{SW}, 
the spectrum of the theory once again consists of a scalar 
($\varphi$) in 
the open and ($\phi$) closed string sector. The closed-\cite{OV} 
and open-string\cite{MAR} 
vertex operators are given by:
\begin{eqnarray}
\label{eq:8}
& & V_o|_{\theta=0}=e^{i(k\cdot{\bar x}+{\bar k}\cdot x)},\ 
V_c|_{\theta={\bar\theta}=0}=e^{i(k\cdot{\bar x}+{\bar k}\cdot
  x)},\nonumber\\
& & V_c^{\rm int}=
\biggl(ik\cdot\partial{\bar x}-i{\bar k}\partial x
-k\cdot{\bar\psi}_R{\bar k}\cdot\psi_R\biggr)
\biggl(ik\cdot{\bar\partial}{\bar x}-i{\bar k}{\bar\partial} x
-k\cdot{\bar\psi}_L{\bar k}\cdot\psi_L\biggr)
e^{i(k\cdot {\bar x}+{\bar k}\cdot x)}
,\nonumber\\
& & V_o^{\rm int}=\biggl(ik\cdot\partial_\tau{\bar x}
-i{\bar  k}\cdot{\partial}_\tau x
-k\cdot({\bar\psi}_L+{\bar\psi}_R){\bar
  k}\cdot(\psi_L+\psi_R)\biggr)
e^{i(k\cdot{\bar x}+{\bar k}\cdot x)},
\end{eqnarray}
where $V^{\rm int}_{c,o}$ are the closed- and open-string vertex
operators that have been integrated w.r.t. their fermionic
supercoordinates. Following \cite{ADEMO}, $\theta_L$ is set equal
to $\theta_R$ for $V^{\rm int}_o$. Also, the bosonic component $x^i,\ 
x^{\bar i}$ denote $(x,y)$ 
and $({\bar x}, {\bar y})$  which originate from 
(anti-)chiral and (anti-)twisted chiral fields of $N=2$. 

The two-point function for both bosons and fermions appearing in
(\ref{eq:8}) can be written 
together using the superspace 2-point function in $N=1$ notation
of \cite{ITY}
(with $\alpha^\prime={1\over{2\pi}}$):
\begin{eqnarray}
\label{eq:6}
& & 
\langle X^i(Z_1,{\bar Z}_1) X^{\bar j}(Z_2,{\bar Z}_2)\rangle
=-g^{i{\bar j}}
ln[(z_1-z_2-\theta_1^L\theta_2^L)({\bar z}_1-{\bar
  z}_2-\theta_1^R\theta_2^R)]\nonumber\\
& & +(g^{i{\bar j}}-2G^{i{\bar j}})ln[(z_1-{\bar
  z}_2-\theta_1^L\theta_2^R)
({\bar z}_1-z_2-\theta_1^R\theta_2^L)]\nonumber\\
& & -2\Theta^{i{\bar j}}
ln\biggl[{z_1-{\bar z}_2-\theta_1^L\theta_2^R\over{\bar z}_1
-z_2-\theta_1^R\theta_2^L}\biggr].\nonumber\\
& & 
\end{eqnarray}
The indices $i, {\bar j}$ run over $1,2$ and ${\bar 1}, {\bar 2}$
respectively. The open string metric $G^{i{\bar j}}$ and the
noncommutativity parameter $\Theta^{i{\bar j}}$ can be expressed in 
terms of $g^{i{\bar j}}$ and $B^{i{\bar j}}$ as in \cite{SW}. For our case, 
$g^{i{\bar j}}$ denotes the flat closed string metric and 
$B^{i{\bar j}}$, constant antisymmetric background of `magnetic' type.

Now, using the above results, we calculate various string amplitudes
in the open and mixed sectors. In the closed sector the results of 
\cite{OV,MAR} are still valid as closed strings have no boundary,
and hence are insensitive to the addition of boundary terms to the
world-sheet action. 
In the open- and mixed-string sectors, the super-M\"{o}bius
transformations allow two complex  
fermionic supercoordinates to be set to zero, and three real bosonic
coordinates to be fixed to any arbitrary value.

(I) \underline{Open sector}

The 3-point function using obvious notations is given by:
\begin{eqnarray}
\label{eq:9}
& & A_{ooo}(B\neq0)=\langle V_o|_{\theta=0}(0) V_o^{\rm int}(1)
V_o(\infty)|_{\theta=0}\rangle
\nonumber\\
& & = e^{{i\over2}({\bar k_1}\Theta k_2-{\bar k}_2\Theta
  k_1)}A_{ooo}(B=0),
\end{eqnarray}
where
$A_{ooo}(B=0)=c_{12}\equiv k_1 G^{-1}{\bar k}_2 -k_2G^{-1}{\bar
  k}_1$. 
Now, as in \cite{MAR}, one has to
impose Bose symmetry on $A_{ooo}$ in 
(\ref{eq:9}). Unlike \cite{MAR}, for the noncommutative case,
one can have an ``isoscalar'' as well as an ``isovector'' component
of the amplitude:
\begin{equation}
\label{eq:isoscvec1}
A_{ooo}=A_{ooo}^{\rm S}+A_{ooo}^{\rm AS\ abc},
\end{equation}
where $A_{ooo}^{\rm S}$ is the isoscalar part of the 
amplitude that is symmetric under the interchange of the
momentum labels of particles 1 and 2, and
$A_{ooo}^{\rm AS\ abc}$
is the isovector part of the amplitude that is
antisymmetric under the interchange of the momentum and group 
labels separately, but is symmetric under the simultaneous interchange
of both types of labels. Denoting
$K_{12}\equiv{1\over2}({\bar k_1}\Theta k_2-{\bar k}_2\Theta k_1)$
one sees that 
\begin{equation}
\label{eq:isoscvec2}
A_{ooo}^{\rm S}=c_{12}sin(K_{12}),\>\>\>\>
A_{ooo}^{\rm AS\ abc}=c_{12}cos(K_{12})f^{abc}.
\end{equation}

Similarly, the 4-point function is given by:
\begin{eqnarray}
\label{eq:10}
& & A_{oooo}(B\neq0)=\int_0^1\langle V_0|_{\theta=0}(0) V^{\rm
  int}_o(x) V^{\rm int}_o(1)V_o|_{\theta=0}(\infty)\rangle\nonumber\\
& & = e^{i(K_{12}+K_{23}+K_{13})}A_{oooo}(B=0),
\end{eqnarray}
where $A_{oooo}(B=0)=F{\Gamma(1-2s)\Gamma(1-2t)\over\Gamma(2u)}$ as in 
\cite{MAR}. The $\Theta$-dependent phase factor in
(\ref{eq:10}) matches with the phase factor in equation (2.11) of \cite{SW}.
The null kinematic factor $F$ is that of \cite{OV}
with the difference that the  open-string metric is used for 
contracting momenta in $s, t$, $u$ as well as $c_{ab}$'s.

Moreover, since in 
the purely open string
sector, the $\Theta$ dependence of the amplitude enters via a phase
factor, one sees that the above result for the
noncommutative 3- and 4-point functions readily generalizes to the
noncommutative $n$-point function, implying that, like the claim for
the commutative $N=2$ theory, all $n$-point functions with $n\geq4$
also vanish. Hence, the noncommutative $N=2$ theory is ``topological'' 
in the closed- and open-string sectors.

We now analyze in some detail the implications of the above
modifications to the field theory of open string scalars.
Using (\ref{eq:isoscvec1}) and (\ref{eq:isoscvec2}),  one can
evaluate the field theory (FT) amplitude 
$A_{oooo\ FT}$ which consists of contributions from 
two 3-point functions ($A_{ooo\ FT}$'s)
 as well as a contact vertex $V_{oooo\ FT}$
(whose form is determined from the requirement that $A_{oooo\ FT}$
like $A_{oooo}$ of string theory, vanishes). One can verify that
for $A_{ooo\ FT}$ corresponding to $A^S_{ooo}$ in (\ref{eq:isoscvec2}),
there are no poles in $A_{oooo\ FT}$. This can be seen by adding
contributions to the 4-point amplitudes from $s$, $t$ and $u$-channels
as:
\begin{equation}
\label{eq:Ampl}
A^S_{oooo FT} = A sin(K_{12}) sin(K_{34}) + B sin(K_{23}) sin(K_{41})
+ C sin(K_{31}) sin (K_{24}),
\end{equation}
where 
\begin{equation}
\label{eq:ABCdefs}
A={c_{12}c_{34}\over s},\ B={c_{23}c_{41}\over t}=u-A,\
C={c_{31}c_{24}\over u}=t+A. 
\end{equation}
Then by eliminating $k_4$ using momentum conservation, it is noticed
that pole part of the amplitude above cancels for $A_{ooo}^{S}$ in 
equation (\ref{eq:Ampl}). In other words, $sin(K_{ab})$ in
 $A^S_{oooo FT}$ acts as a structure constant. To generalize this
 result further, one can consider a more general 3-point function 
\begin{equation}
\label{eq:oood}
A^{S\ abc}_{ooo}=c_{12}sin(K_{12})d^{abc},
\end{equation}
 with $d^{abc}$ being
symmetric structure constants. Vanishing of poles in $A^S_{oooo\ FT}$ 
then implies a strong condition on $d^{abc}$'s
leading to multiple copies of abelian noncommutative FT's 
mentioned in (\ref{eq:openaction}) below.

For $A_{ooo}^{AS}$, on the other hand, we get
a constraint on $f^{abc}$:
\begin{eqnarray}
\label{eq:ASconstr}
& & cos(K_{12})sin(K_{31})sin(K_{32})f^{abx}f^{xbd}
-cos(K_{23})sin(K_{21})sin(K_{31})f^{bcx}f^{xda}\nonumber\\
& & +cos(K_{31})sin(K_{21})sin(K_{23}) f^{cax}f^{xbd}=0.
\end{eqnarray}
One sees that the above constraint can not be satisfied by any 
classical group. Perhaps it may be satisfied for some quantum 
group. One now observes that 
$U(N)$ gauge groups can be obtained from 3-point string vertex,
eqn.(\ref{eq:9}), by considering mixed (isoscalar-isovector)
vertices.
In particular, for $U(2)$, after imposing Bose
symmetry on two of the external legs in eqn.(\ref{eq:9}), the 
corresponding isoscalar-isovector field theory vertex is given as: 
\begin{equation}
A^{a b}_{Mixed} = c_{12}sin(K_{12})\delta^{ab}.
\end{equation}
Then it can once again be shown that the poles in the isovector 4-point 
amplitude, obtained by sewing together two 3-point vertices with 
isoscalar and isovector internal states, cancel\footnote{We thank the
referee to point this out.}. Higher rank groups 
can also be incorporated by including 3-point vertex in 
eqn.(\ref{eq:oood}) (See for example \cite{lech}). 

We now write down the FT corresponding to $A^S_{ooo\ FT}$ mentioned 
before, as well as the contact vertex appearing in equation (\ref{eq:Ampl}),
after using equation (\ref{eq:ABCdefs}) whose explicit form is
\begin{equation}
\label{eq:contactV}
V^{int}_{oooo} = u sin(K_{23}) sin(K_{41}) + t sin(K_{31})
sin(K_{24}).
\end{equation}
One then obtains 
the field theory action corresponding to $A^S_{ooo}$ and
$V^{int}_{oooo}$ up to terms quartic in $\varphi$: 
\begin{eqnarray}
\label{eq:openaction}
& & 
{\cal L}_{FT}=G^{{\bar i}j}
\biggl[{1\over2}{\bar\partial}_{\bar i}\varphi * \partial_j\varphi
+{i\over3}[{\bar\partial}_{\bar i}\varphi,\partial_j\varphi]_*
*\varphi-{1\over12}{\bar\partial}_{\bar i}\varphi*
[[\partial_j\varphi,\varphi]_*,\varphi]_*\biggr],\nonumber\\
& & 
\end{eqnarray}
where $[\xi,\eta]_*\equiv \xi*\eta-\eta*\xi$. The $*$ product is defined
(in momentum space) as:
\begin{equation}
\label{eq:Moydef}
e^{ik_1\cdot x}*e^{ik_2\cdot x}=e^{iK_{12}}e^{i(k_1+k_2)\cdot x}.
\end{equation}
with $k\cdot x \equiv k\cdot\bar{x} + \bar{k}\cdot x$. 
A generalization of the above noncommutative abelian field theory action
to $U(N)$ case is straightforward. 
The Moyal deformation of self dual Yang Mills (and gravity) were also
considered in \cite{PLEBA}. We now study the mixed sector of 
the noncommutative $N$=2 theory.

(II) \underline{Mixed sector}

(a) $A_{ooc}$: We now  show, in $N=2$ context, 
that a mixed  amplitude with 
two open and one closed string vertices generates a field
theory with a generalized *-product\cite{GAROUSI,MEHEN} 
(the $\Theta=0$ limit of which reduces to the result of
\cite{MAR}). 

Following
\cite{MAR}, for the purpose of setting the limits of
integration, it is convenient to fix the bosonic coordinate of one
of the two $V_o$'s (to 0) and that of $V_c$ to $z=x+iy$. Then 
\begin{eqnarray}
\label{eq:11}
& & A_{ooc}=\int_{-\infty}^\infty db \langle V_o|_{\theta=0}(b)
V^{\rm int}_o(z=x+iy)
V_o|_{\theta=0}(\tau\rightarrow\infty)\rangle\nonumber\\
& & = e^{{i\over2}({\bar k}_b\Theta k_\tau-{\bar k}_\tau\Theta k_b)}\times
\nonumber\\ 
& & \times {y\over{4\pi^2}} \int_{-\infty}^\infty db \biggl(c_{b\tau}^2-
(k_\tau\Theta{\bar k}_b+k_b\Theta {\bar k}_\tau)^2\biggr)
{e^{-{1\over{2\pi}}(k_b\Theta{\bar k}_\tau-k_\tau
\Theta{\bar  k}_b)ln
\biggl({b-(x-iy)\over{b-(x+iy)}}\biggr)}\over{([b-x]^2 +
y^2)}}\nonumber\\
& & = 
{e^{{i\over2}({\bar k}_b\Theta k_\tau-{\bar k}_\tau\Theta k_b)}\over{4\pi^2}}
\biggl(c_{b\tau}^2-
(k_\tau\Theta{\bar k}_b+k_b\Theta {\bar k}_\tau)^2\biggr)
\pi e^{-i{(k_b\Theta{\bar k}_\tau-k_\tau
\Theta{\bar  k}_b)\over2}}{sin\biggl(
{k_b\Theta{\bar k}_\tau-k_\tau\Theta{\bar k}_b\over2}
\biggr)\over{(k_b\Theta{\bar k}_\tau-k_\tau\Theta{\bar k}_b)}}\nonumber\\
& & ={\pi\over2} c^L_{b\tau}c^R_{b\tau}{sin(K_{b\tau})\over{K_{b\tau}}},
\end{eqnarray}
where 
$c^{L,R}_{ab}\equiv{1\over{2\pi}}\biggl(k_aG^{-1}{\bar k}_b-
k_bG^{-1}{\bar k}_a \mp k_a\Theta{\bar k}_b
\pm k_b\Theta{\bar k}_a\biggr)$, the upper and lower signs
corresponding to $L$ and $R$ respectively. The $b$ integral above was
done using Mathematica and is also given in appendix A of \cite{GARMYRS}.

Before commenting on the topological nature of this amplitude, we
now  write down the corresponding interaction term in the 
FT action which is
given in terms of a generalized *-product\cite{GAROUSI,MEHEN}: 
\begin{equation}
\label{eq:mixedaction}
{\cal L}_{ooc} = \phi
(\partial_i\partial_{j}
\varphi
*^\prime
{\bar\partial}_{\bar i}{\bar\partial}_{{\bar j}}\varphi
-\partial_i{\bar\partial}_{{\bar i}}
\varphi *^\prime
\partial_{j}
{\bar\partial}_{{\bar j}}\varphi)
(G^{-1}-\Theta)^{i{\bar j}}(G^{-1}+\Theta)^{j{\bar i}},
\end{equation}
where we have made use of 
\begin{equation}
\label{eq:Pr*def}
e^{ik_1\cdot x} *^\prime e^{ik_2\cdot x}=
{sin(K_{12})\over{K_{12}}}e^{i(k_1+k_2)\cdot x}.
\end{equation}
in arriving at (\ref{eq:mixedaction}) from (\ref{eq:11}). 

Now, to interpret
$A_{ooc}$ as a topological theory in the sense of \cite{OV}, one now
sees that by expanding ${sin(K_{12})\over{K_{12}}}$
in a power series in $\Theta$, an infinite number of
($\Theta$ or equivalently $B$-dependent) terms are generated 
at the 3-point level in the mixed sector. As the
radius of convergence of the
${sin x\over x}$ expansion is infinite, this implies that {\it after
expansion},
$A_{ooc}$ can be interpreted as an infinite series of
local interactions between the closed and open string scalars. 

(b) $A_{oooc}$:
Like
\cite{MAR}, we set the bosonic coordinates of the three $V_o$'s
at $0, 1, \infty$ and the fermionic coordinates of the first and third
$V_o$'s to zero. We will hence require to integrate over 
the bosonic coordinates of $V_c^{\rm int}(z,{\bar z})$. Defining
$t^{L,R}\equiv{1\over{2\pi}} 
\biggl( k_1G^{-1}{\bar k}_4+k_4G^{-1}{\bar k}_1
\mp k_1\Theta{\bar k}_4\mp k_4\Theta{\bar k}_1\biggr)$, 
$u^{L,R}
\equiv{1\over{2\pi}} \biggl( k_1G^{-1}{\bar k}_3+k_3G^{-1}{\bar k}_1
\mp k_1\Theta{\bar k}_3\mp k_3\Theta{\bar k}_1\biggr)$,
one gets:
\begin{eqnarray}
\label{eq:13}
& & A_{oooc}=\int\int_{\rm UHP} dz d{\bar z}
\langle V_o|_{\theta=0}(0)V^{\rm int}_o(1)V_o|_{\theta=0}(\infty) 
V^{\rm int}_c(z,{\bar z})\rangle\nonumber\\
& & \sim \int\int_{\rm UHP} dz d{\bar z}
z^{t^L}{\bar z}^{t^R}(1-z)^{u^L}(1-{\bar z})^{u^R}\nonumber\\
& & \times
\Biggl(\biggl[-{c^L_{14}\over z}+{c^L_{24}\over(1-z)}\biggr]\biggl[
-{c^R_{14}\over {\bar z}}+{c^R_{24}\over(1-{\bar
    z})}\biggr]\biggl[-2c_{12}
+{c^L_{24}\over(1-z)}+{c^R_{24}\over(1-{\bar z})}\biggr]\nonumber\\
& & +\biggl[-{c^L_{14}\over z}+{c^L_{24}\over(1-z)}\biggr]\biggl[{-u^R
+(u^R)^2-(c^R_{24})^2\over(1-{\bar z})^2}\biggr] \nonumber\\
& &  +\biggl[-{c^R_{14}\over {\bar z}}+{c^R_{24}\over(1-{\bar z})}
\biggr]\biggl[{-u^L
+(u^L)^2-(c^L_{24})^2
\over(1-z)^2}\biggr]\Biggr),
\end{eqnarray}
which gives the $A_{oooc}$ amplitude of \cite{MAR} for $\Theta=0$.
However, because 
of the lack of $z\rightarrow{\bar z}$ symmetry in the presence 
of $B$, unlike \cite{MAR}, one can not enlarge the domain of
integration from the upper half complex plane to the entire complex
plane

To simplify the algebra,
one notices that in the extreme noncommutative case
($\Theta\rightarrow\infty$), which has 
been a subject of recent interest\cite{SOLITONS},   
$t^L=-t^R = \tilde{t}$, $u^L=-u^R=\tilde{u}$ and
$c^L_{ab}=-c^R_{ab}=\tilde{c}_{ab}$. 
We will now consider only this case in 
the paper. One can then verify that the coefficient
of the leading term vanishes when 
one makes use of the fact that any finite powers of $z, \bar{z}, 
(1-z), (1-\bar{z})$ can be dropped with respect to the ones 
containing $t^{L, R}, u^{L, R}$. This suggests that 
4-point amplitude of (\ref{eq:13}) is possibly zero.
We now argue 
that each of the integrals appearing in eqn.(\ref{eq:13}) 
is  in fact zero for generic non-integral $\tilde{t}$
and $\tilde{u}$, positive integral 
values of $\tilde{t}$ or $\tilde{u}$, as well as negative integral
values of $\tilde{u},\tilde{t}<-4$. We hence have to evaluate
integrals of the type
\begin{equation}
\label{eq:14}
\int\int dz d{\bar z}
z^a(1-z)^b{\bar z}^{c}(1-{\bar z})^{d}
\end{equation}
over the UHP.  The above integral is similar, though not 
identical to the ones that are evaluated in \cite{GARMYRS}. 

Using the Stokes theorem in the complex plane,
the integral of
(\ref{eq:14}) gets mapped to 
\begin{equation}
\label{eq:16}
\biggl(\int_{\rm Im z=0\ axis} + \int_{C_R})
{{\bar z}^c(1-{\bar z})^d
z^{1+a}\over{1+a}}\ _2F_1(1+a,-b,a+2;z)
,
\end{equation}
where $C_R$ is a semicircular contour in the upper half plane whose
radius is taken to infinity eventually. Now, we use the integral 
representation of $\ _2F_1(1+a,-b,a+2;z)$ as given in equation 
{\bf 15.3.1} of \cite{ABST} valid for $a+2>-b>0,\ |Arg(1-z)|<\pi$.
The term $c_{12}$ in equation (\ref{eq:13}) drops 
out as compared to terms consisting of $c^L_{24}$ or $c^R_{24}$.
From (\ref{eq:13}) one will pick up an $R^{-3}$ from each of the terms.
So,
when evaluating $\int_{C_R}$, one gets
$\lim_{R\rightarrow\infty}\int_C\sim R^{c+d-2}$.
Hence, one takes $c+d<2$ for the purpose of evaluation of the
integral so that $\lim_{R\rightarrow\infty}\int_C=0$,
 keeping in mind that the answer that one gets can
be analytically continued to $c+d\geq2$ domain, and those  $a, b$ not
satisfying the above constraints.
Hence, one is left with the integral over the real axis 
which is:
\begin{equation}
\label{eq:19}
\int_{-\infty}^\infty dx\ x^{a+c+1}(1-x)^d\ _2F_1(1+a,-b,a+2;x).
\end{equation}
If $b\in{\bf Z}^+$, then $\ _2F_1(1+a,-b,a+2;x)$ can be expanded as
a finite series, and using that $\int_{-\infty}^\infty dx
x^\alpha(1-x)^\beta$ is proportional to $sin(\pi\alpha)$ (for more exact
expression, see  (\ref{eq:24})), one sees that (\ref{eq:19}) vanishes.
For $b\in{\bf Z}^-$ and $b<-4$, one performs the ${\bar z}$ integration first,
and the argument for $b\in{\bf Z}^+$ follows here as well. For
$a\in{\bf Z}$, we conformally map the UHP to the LHP by $z\rightarrow (1-z)$.
Then the argument for $b\in{\bf Z}$ can be repeated here. So, 
(\ref{eq:19}) vanishes if at least one of $a$ or $b$ is a positive
integer or a negative integer less than -4. 
We now argue that (\ref{eq:19}) vanishes for $a,b\not\in{\bf Z}$
as well.

The identity {\bf 15.3.6} of \cite{ABST}
which is valid for $|Arg(1-z)|<\pi$, is now used. 
The integrand in (\ref{eq:19}) is analytic in the entire complex plane
except along the branch cuts from $x=0$ to $x=1$ and along $x>1$. Hence,
one can deform the contour along
the real axis to a contour $C$ (which is  equivalent to putting 
$z=x+i\epsilon$ for $x>1$) of Fig. 1(a). Along $C$, 
$|Arg(1-z)|<\pi$ is satisfied for the entire contour. After applying
{\bf 15.3.6} (of \cite{ABST})
 to (\ref{eq:19}), one deforms $C$ of Fig. 1(a) back
to the contour of Fig. 1(b).

\begin{figure}[htbp]
\vskip -1.5 in
\centerline{\mbox{\psfig{file=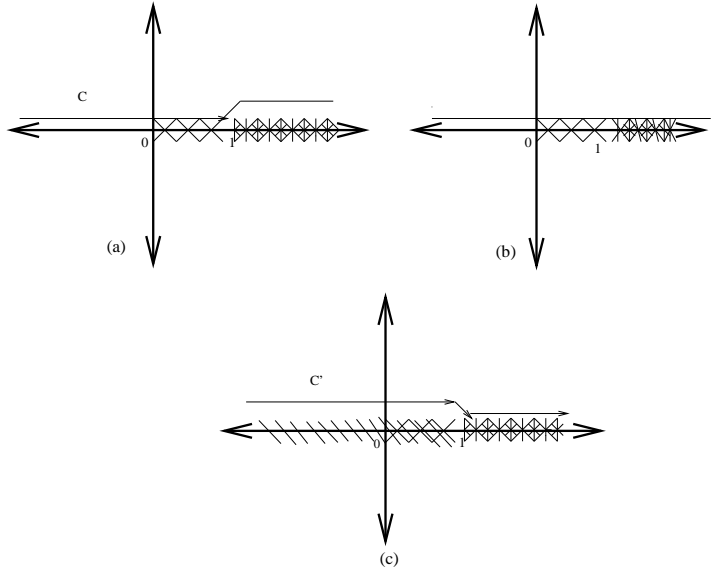,width=0.7\textwidth}}}
\vskip -1 in
\caption{Deformation of contour of integration; 
there are three branch cuts: $x\rightarrow-\infty$ to $x=1$, $x=0$ to
$x=1$, and $x=1$ to $x\rightarrow\infty$}
\end{figure}

Now,  the Mellin-Barnes contour integral representation of the
hypergeometric function $\ _2F_1(\alpha,\beta,\gamma;(1-z))$ as given 
in {\bf 15.3.2} of \cite{ABST} valid for  $|Arg[-(1-z)]|<\pi$, is used.
Like before, one uses the analyticity property of the
integrand of (\ref{eq:19}), and deforms the contour of Fig. 1(b) to
$C'$ (which is equivalent to setting $z=x+i\epsilon$ for $x<1$) 
of Fig. 1(c). Along $C'$, $|Arg[-(1-z)]|<\pi$ is satisfied for the
entire contour. 
Finally, we deform $C'$ back to Fig. 1(b).
We will thus have:
\begin{equation}
\label{eq:22}
\int_{-\infty}^\infty dx x^{a+c+1}(1-x)^{d+s+\lambda},
\end{equation}
where $\lambda=0$ or $1+b$. The above integral, after evaluation, 
has a form:
\begin{equation}
\label{eq:24}
\int_{-i\infty}^{i\infty}ds
{sin(\pi[a+c+1])sin(\pi[d+s+\lambda])\over{sin(\pi[a+c+d+s+\lambda+1])}}
{\Gamma(a+c+2)\Gamma(d+s+\lambda+1)\over{\Gamma(a+c+d+s+\lambda+3)}},
\end{equation}
with certain restrictions on $a, c, d, \lambda$ which can then 
be removed by analytic continuation.
In the large $B$ limit $a+c\in {\bf Z}$, 
hence $sin(\pi[a+c+1])=0$ implying that (\ref{eq:19}) vanishes. 
As a result, $A_{oooc}(B\rightarrow\infty)=0$ in all the cases
discussed above.

We end by pointing out that 
it will be interesting to examine the applications of $N=2$ 
noncommutativity to M(atrix) and F theories.

\end{document}